\documentclass{appolb}

\usepackage[english]{babel}
\usepackage[utf8]{inputenc}
\usepackage{amsthm,amssymb,amsmath}  
\usepackage{mathtools}
\usepackage{tensor}
\usepackage{graphicx}
\usepackage{float}
\usepackage{bpchem}
\usepackage{bbm}
\usepackage[citecolor=black]{hyperref}
\usepackage{booktabs}
\usepackage{color}
\definecolor{darkblue}{rgb}{0,0,0.3}
\hypersetup{colorlinks=true, linkcolor=darkblue,urlcolor=darkblue}
\usepackage{fancyhdr}
\usepackage{cite}

\usepackage{graphicx}

\begin{document}
\title{The sunrise integral around two and four space-time dimensions in terms of elliptic polylogarithms%
}
\author{Luise Adams$^1$\thanks{presented at 'Matter To The Deepest' Conference, 14$^{th}$ to 18$^{th}$ September 2015, Ustro\'n, Poland}, 
Christian Bogner$^2$, Stefan Weinzierl$^1$
\address{$^1$PRISMA Cluster of Excellence, Institut f\"ur Physik, Johannes Gutenberg Universit\"at Mainz, D - 55099 Mainz, Germany}
\\
\address{$^2$ Institut f\"ur Physik, Humboldt-Universit\"at zu Berlin, D - 10099 Berlin, Germany}
}
\maketitle
\begin{abstract}
In this talk we discuss the solution for the sunrise integral around two and four space-time dimensions in terms of a generalised elliptic version of the multiple polylogarithms. In two space-time dimensions we obtain a sum of three elliptic dilogarithms.
The arguments of the elliptic dilogarithms have a nice geometric interpretation.
In four space-time dimensions the sunrise integral can be expressed with the $\epsilon^0$- and $\epsilon^1$-solution around two dimensions, mass derivatives thereof and simpler terms.
\end{abstract}

\begin{samepage}
\section{Introduction}
The non-vacuum two-loop sunrise integral is the simplest integral in quantum field theory which cannot be solved in terms of multiple polylogarithms.
Because of its fundamental character the sunrise integral received numerous attention 
in the past \cite{Smirnov:2004ym,Davydychev:1997wa,Broadhurst:1987ei,Tarasov:2006nk,Bailey:2007,Kalmykov:2008ge,Bailey:2008ib,Broadhurst:2008mx,Caffo:1998du,Laporta:2004rb,MullerStach:2011ru,MullerStach:2012mp,Adams:2013nia,Adams:2014vja,Adams:2015gva,Bloch:2013tra}.
For the equal mass case an analytic result involving elliptic integrals and elliptic polylogarithms was provided in \cite{Laporta:2004rb,Bloch:2013tra}. 
The results can be generalised to the arbitrary mass case \cite{Adams:2013nia,Adams:2014vja,Adams:2015gva}, which we will discuss in this talk. 

\section{Differential equation for the sunrise integral}
In $D$-dimensional Minkowski space the two-loop sunrise integral reads
\end{samepage}

\begin{align}
& S_{\nu_1 \nu_2 \nu_3} (D,p^2,m_1^2,m_2^2,m_3^2,\mu^2) = \\
& (\mu^2)^{\nu-D} \int \dfrac{d^Dk_1}{\text{i} \pi^{\frac{D}{2}}} \dfrac{d^D k_2}{\text{i} \pi^{\frac{D}{2}}} \dfrac{1}{(-k_1^2 + m_1^2)^{\nu_1} (-k_2^2+m_2^2)^{\nu_2} (-(p-k_1-k_2)^2 + m_2^2)^{\nu_3}} \nonumber
\end{align}
with the three internal masses $m_1, m_2, m_3$. 
In terms of Feynman parameters we obtain for the sunrise integral the expression
\begin{align}
S_{\nu_1 \nu_2 \nu_3} (D,t) = \dfrac{\Gamma(\nu-D)}{\Gamma (\nu_1) \Gamma (\nu_2) \Gamma (\nu_3)} (\mu^2)^{\nu-D} \int\limits_{\sigma} x_1^{\nu_1-1} x_2^{\nu_2-1} x_3^{\nu_3-1} \dfrac{\mathcal{U}^{\nu-\frac{3}{2} D}}{\mathcal{F}^{\nu-D}} \omega
\label{FeynmanParameterIntegral}
\end{align}
with $\nu = \nu_1 + \nu_2 + \nu_3$, the momentum squared $t=p^2$, 
the differential two-form $\omega = x_1\ dx_2 \wedge dx_3 - x_2\ dx_1 \wedge dx_3 + x_3\ dx_1 \wedge dx_2$,
the integration region $\sigma = \lbrace [x_1 : x_2 : x_3] \in \mathbb{P}^2 | x_i \geq 0, i=1,2,3 \rbrace$
and the Feynman graph polynomials $\mathcal{U}$ and $\mathcal{F}$
\begin{align}
\mathcal{U}=x_1 x_2 + x_2 x_3 + x_3 x_1, \qquad \mathcal{F} = -x_1 x_2 x_3 t + (x_1m_1^2 + x_2m_2^2 + x_3m_3^2)\ \mathcal{U}.
 \nonumber
\end{align}
We are interested in the Laurent expansions of $S_{111}(D,t)$ around two and four space-time dimensions. These expansions start as
\begin{align}
S_{111}(2-2\epsilon,t) &= e^{-2\gamma \epsilon}\ [S^{(0)}_{111}(2,t) + \epsilon S^{(1)}_{111}(2,t) + \mathcal{O}(\epsilon^2)], \\
S_{111} (4-2\epsilon,t) &= e^{-2\gamma \epsilon} \left[ \dfrac{1}{\epsilon^2} S^{(-2)}_{111}(4,t) + \dfrac{1}{\epsilon} S^{(1)}_{111}(4,t) + S_{111}^{(0)}(4,t) + \mathcal{O} (\epsilon)\right].
 \nonumber
\end{align}
In the following we are mainly interested in determining $S^{(0)}_{111}(2,t)$ and $S^{(0)}_{111}(4,t)$.\\
In $D$ dimensions the sunrise integral $S_{111}(D,t)$ satisfies an inhomogeneous fourth-order differential equation \cite{Adams:2015gva}:
\begin{align}
\left[ \sum\limits_{i=0}^4 P_i \dfrac{d^i}{dt^i} \right] S_{111}(D,t) = \mu^2 \left[ c_{12} T_{12} + c_{13} T_{13} + c_{23} T_{23} \right]
\label{DGLDDim}
\end{align}
where the $P_i$'s and $c_{ij}$'s are polynomials in $D$, $t$ and the masses and the $T_{ij}$'s are products of simpler tadpole integrals. 
This differential equation can then be expanded for example around $D=2-2\epsilon$.

\section{The sunrise integral around two space-time dimensions}
In two space-time dimensions the fourth-order differential equation for the sunrise integral from eq.(\ref{DGLDDim}) can be simplified to a second-order differential equation with a polynomial $p(t)$ depending on $t$, the masses and logarithms of the masses \cite{MullerStach:2011ru}:
\begin{align}
L_2^{(0)}(2)\ S^{(0)}_{111}(2,t) = \mu^2 p(t).
\end{align}
In $D=2$ the integrand of eq.~(\ref{FeynmanParameterIntegral}) 
simplifies to $\frac{1}{\mathcal{F}}$.
The equation $\mathcal{F}=0$
defines together with the choice of an origin an elliptic curve. A convenient choice for the origin is one of the intersection points of the algebraic variety $\mathcal{F}=0$ and the integration region $\sigma$ where we have three different points $P_1=[1:0:0], P_2=[0:1:0]$ and $P_3=[0:0:1]$. Every chosen origin $P_i$ corresponds to a certain elliptic curve $E_i$ with $i=\lbrace 1,2,3 \rbrace$ which we can transform into the Weierstrass normal form $\hat{E}_i$. \\
In the next step, we compute the periods $\Psi_1, \Psi_2$ of the elliptic curve $\hat{E}_i$ which leads to complete elliptic integrals of the first kind:
\begin{align}
\Psi_1 = \dfrac{4\mu^2}{\sqrt{D}} K(k), \qquad \Psi_2 = \dfrac{4\mu^2 \text{i}}{\sqrt{D}} K(k')
\end{align}
where the (complementary) modulus $k^{(')}$ depends on the roots $e_i$ of the Weierstrass normal form while the polynomial $D$ depends on $t$ and the masses.
The periods $\Psi_1, \Psi_2$ are the homogeneous solutions of the second-order differential equation.\\
After changing the variable from $t$ to the nome $q = e^{2 \pi \text{i} \tau}$ (cf. \cite{Bloch:2013tra}) which is the natural variable on an elliptic curve with the period ratio $\tau=\frac{\Psi_2}{\Psi_1}$ we can write the special inhomogeneous solution in an overseeable form. The aim is now to express this inhomogeneous solution in terms of the homogeneous solutions and a new (elliptic) generalisation of the classical polylogarithms.
The two periods imply a lattice $\Lambda$ which belongs to the torus representation of the elliptic curve $\mathbb{C} \text{\textbackslash} \Lambda$ to which we can transform by the following elliptic integral
\begin{align}
[x:y:1] \rightarrow \hat{z} = \dfrac{1}{\Psi_1} \int\limits_x^{\infty} \dfrac{1}{\sqrt{4(x-e_1)(x-e_2)(x-e_3)}}.
\end{align}
This function maps the image of the intersection point $P_j \in E_k$ on the curve $\hat{E}_k$ abbreviated by $Q_{j,k}$ onto the point
\begin{align}
Q_{j,k} \rightarrow \hat{z}_i = \dfrac{1}{2} \dfrac{F(u_i,k)}{K(k)}
\end{align}
with the incomplete elliptic integral of the first kind $F(z,x)$.\\
As last step we can compute the Jacobi uniformization $\mathbb{C}^* / q^{2\mathbb{Z}}$ of the elliptic curve via the complex exponential $\hat{z} \rightarrow w = e^{2\pi \text{i} \hat{z}}$ mapping the points $\hat{z}_i$ to
\begin{align}
\hat{z}_i \rightarrow w_i = \exp \left[ \text{i} \pi \dfrac{F(u_i,k)}{K(k)} \right].
\end{align}
It turns out that with the following definition of elliptic polylogarithms
\begin{align}
\label{defELi}
\text{ELi}_{n;m} (x;y;q) & = \sum\limits_{j=1}^{\infty} \sum\limits_{k=1}^{\infty} \dfrac{x^j}{j^n} \dfrac{y^k}{k^m} q^{jk}
 \\
\text{E}_{n;m} (x;y;q) & = 
 \dfrac{c_{n+m}}{\text{i}} \left[ \left( \dfrac{1}{2} \text{Li}_n(x) + \text{ELi}_{n;m} (x;y;q) \right) 
 \right. \nonumber \\
 & \left. - s_{n+m} \left(\dfrac{1}{2} \text{Li}_n (x^{-1}) + \text{ELi}_{n;m} (x^{-1}; y^{-1};q) \right)\right], \nonumber
\end{align}
(with $c_n=1$, $s_n=1$ for $n$ even and $c_n=i$ and $s_n=-1$ for $n$ odd)
the full result for the sunrise integral $S^{(0)}_{111}(2,t)$ in two space-time dimensions with arbitrary masses can be expressed as a sum of three elliptic dilogarithms
\begin{align}
S^{(0)}_{111}(2,t) = \dfrac{\Psi_1(q)}{\pi} 
 \sum\limits_{i=1}^3 \text{E}_{2;0} (w_i;-1;-q) 
\end{align}
with the numbers $w_i$ as arguments \cite{Adams:2014vja}.

\section{The sunrise integral around four space-time dimensions}
The sunrise integral around four space-time dimensions can be expressed in the basis \cite{Caffo:1998du}
\begin{align}
\left\{ \mu^2 S_{111}(D,t),\ \mu^2 \dfrac{\partial}{\partial m_i^2} S_{111}(D,t) = -S_{1+\delta_{i,1}\ 1+\delta_{i,2}\ 1+\delta_{i,3}} (D,t) \right\}
\end{align}
with the help of Tarasov's dimensional shift relations \cite{Tarasov:1996br,Tarasov:1997kx}:
\begin{align}
S^{(0)}_{111}(4,t) = \tilde{L}_3^{(-1)}(2)\ S^{(1)}_{111}(2,t) + \tilde{L}_3^{(0)}(2)\ S^{(0)}_{111}(2,t) + \tilde{R}^{(0)}
\end{align}
where $\tilde{L}_3^{(k)}(2) = C_0^{(k)} + \sum\limits_{i=1}^{3} C_i^{(k)} m_i^2 \frac{\partial}{\partial m_i^2}$ and $\tilde{R}^{(0)}$ is a remainder depending on $t$ and simpler mass terms.
The $\epsilon^1$-solution $S^{(1)}_{111}(2,t)$ involves apart from the functions of eq.~(\ref{defELi}) one new function
$E_{n_1,n_2;m_1,m_2;2o}(x_1,x_2;y_1,y_2;q)$ defined in \cite{Adams:2015gva}.
This function can be viewed as a generalisation of multiple polylogarithms of depth greater than one.
If we define the weight of $E_{n;m}$ by $w=n+m$ and the weight of $E_{n_1,n_2;m_1,m_2;2o}$ by $w=n_1+n_2+m_1+m_2+2o$,
it is worth noting that the ${\mathcal O}(\epsilon^1)$-part is not of uniform weight.
The ${\mathcal O}(\epsilon^1)$-part
of the sunrise integral around two space-time dimensions contains terms of weight three and four.\\
To understand this issue one can consider the simpler equal mass case for $S^{(1)}_{111}(2,t)$ where one is concerned with an inhomogeneous second-order differential equation for $S^{(1)}_{111}(2,t)$. With a suitable ansatz for $S^{(1)}_{111}(2,t)$ involving a part proportional to $S^{(0)}_{111}(2,t)$ and a remainder one obtains a much simpler differential equation. Both parts of the ansatz contain $\log(-q)$-terms which cancel out in the end, but as a remainder from integrating the $\log(-q)$-term from the second part when solving the differential equation we obtain the weight four terms.

\bibliography{/home/stefanw/notes/biblio}
\bibliographystyle{/home/stefanw/latex-style/h-physrev5}

\end{document}